\documentclass{article}
\usepackage{spconf,amsmath,graphicx}

\usepackage{xcolor}
\usepackage{url}

\title{TransMask: A Compact and Fast Speech Separation Model Based on Transformer}
\name{Zining Zhang$^{1,2}$, Bingsheng He$^2$, Zhenjie Zhang$^1$}
\address{
  $^1$PVoice Technology Singapore\\
  $^2$School of Computing, National University of Singapore}
\begin{document}
%
\maketitle
\begin{abstract}

Speech separation is an important problem in speech processing, which targets to separate and generate clean speech from a mixed audio containing speech from different speakers.
Empowered by the deep learning technologies over sequence-to-sequence domain, recent neural speech separation models are now capable of generating highly clean speech audios. 
To make these models more practical by reducing the model size and inference time while maintaining high separation quality, we propose a new transformer-based speech separation approach, called TransMask.
By fully unleashing the power of self-attention on long-term dependency reception, we demonstrate the size of TransMask is more than 60\% smaller and the inference is more than 2 times faster than state-of-the-art solutions.
TransMask fully utilizes the parallelism during inference, and achieves nearly linear inference time within reasonable input audio lengths.
It also outperforms existing solutions on output speech audio quality, achieving SDR above 16 over Librimix benchmark.


\end{abstract}
\section{Introduction}
\label{sec:intro}


It is usually difficult to separate clean audio speeches from the audios in the real world.
In practice, speech processing systems are supposed to handle noisy audio when background musics or even speeches from different speakers are present in the audio clip.
It is therefore crucial to separate the speech from different speakers, before proceeding to further processing and analysis, such as ASR (automatic speech recognition).

With the explosive development of deep learning, recent neural speech separation models, such as TasNet~\cite{luo2018tasnet}, Conv-TasNet~\cite{luo2019conv}, deep CASA~\cite{liu2019divide} and DPRNN~\cite{luo2020dual}, have achieved significant quality improvement over traditional approaches.
The common strategy used in these approaches is to mask the time or frequency domain, with the masks coming from a deep neural network over the audio signals.
The clean audios are then generated by reconstructing the signals based on the masks corresponding to individual speakers.
As two common types of neural network, both Convolutional Neural Network (CNN) and Recurrent Neural Network (RNN) are already employed in the generation of the masks. Recently, self-attention models, such as Transformer and its variants~\cite{vaswani2017attention, beltagy2020longformer, child2019generating}, are performing particularly well in sequence-to-sequence domains, in applications such as neural translation and speech recognition, because of the power of self-attention structure on maintaining long-term dependency.
Such advantages are believed to be beneficial to speech separation tasks as well, since the common challenges in speech separation, such as channel swap, could be dissolved when long-term dependency is well captured. 
 
Moreover, existing RNN-based speech separation models do not scale well with audio length in model training and inference.
The theoretical bound of inference time, for example, is linear to the length of the audio representations.
CNN-based speech separation models are faster, while the output quality of these models are outperformed by RNN-based models because of their limited receptive fields.
Transformers, on the other hand, have the potential to overcome the limitations of both types of models, when 1) self-attention naturally has the maximal receptive field; and 2) the inference can be easily paralleled. 
This means that the cost of sequential operations is nearly constant with sufficient resources,
and the quality of Transformer-based models is expected to be comparable or even better than RNN models.
 
However, as self-attention does not pose auto-regressive regularization as RNNs do, it may suffer from the lack of short-term dependencies.
Therefore, it remains difficult to design a Transformer-based speech separation model meeting all the expectations above.
Some attempts of transformer-based speech separations models~\cite{chen2020dual, subakan2020attention, chen2020continuous} try to solve the problem by either introducing RNN on every transformer layers or using large transformer model which sacrifice the efficiency.
This motivates us to develop a compact, fast, yet effective speech separation model
 with the help of the following designs: \textit{STRNN}, \textit{Sandwich-Norm Transformer Layer} and \textit{Dual-Temporal Convolutional Encoding}.
As a summary, the core contributions of this paper include:
\begin{enumerate}
    \item We propose a new model, called TransMask, incurring nearly constant inference time cost;
    \item We design TransMask to be the smallest deep learning model for speech separation in the literature, 60\% smaller than state-of-the-art solution;
    \item We evaluate TransMask on LibriMix benchmark and demonstrate outstanding SDR performance.
\end{enumerate}



\section{Related work}
\label{sec:relatedbg}

The speaker permutation problem is a significant problem in speech separation.
Based on the solution to this problem, we categorize the speech separation neural models into two categories: deep clustering and permutation invariant training (PIT).
We focus on the prevailing PIT-based models in this paper.


PIT is to calculate losses over all permutations of the outputs for different speakers, and uses the smallest one as the objective for optimization.
There are frame level PIT (tPIT) and utterance level PIT (uPIT)~\cite{kolbaek2017multitalker}. uPIT is prevailing since tPIT is prone to the problem of channel swap.
TasNet~\cite{luo2018tasnet} is one of the successful models using uPIT. It deploys stacked LSTM as the separation module.
Conv-TasNet~\cite{luo2019conv} adopts a similar architecture, but it makes use of Temporal Convolutional Net (TCN) instead of RNN.
It achieves better performance by making the neural network deeper with the help of TCN.
DPRNN~\cite{luo2020dual} uses RNN modules, but with a dual-path procedure, which helps model to achieve better performance than Conv-TasNet.


Most of the existing studies of speech separation is based on the time-frequency representation of the original audios. Instead, TasNet model and its variants apply a trainable encoder and decoder directly over time domain.
The encoder produces a 2D representation from the temporal sequences, and the decoder convert this time-frequency-like representation back to temporal sequences.
The separation module operates on this 2D representation and produces the masks of clean sources.
 
\begin{figure*}[t]
    \centering
    \includegraphics[width=7in]{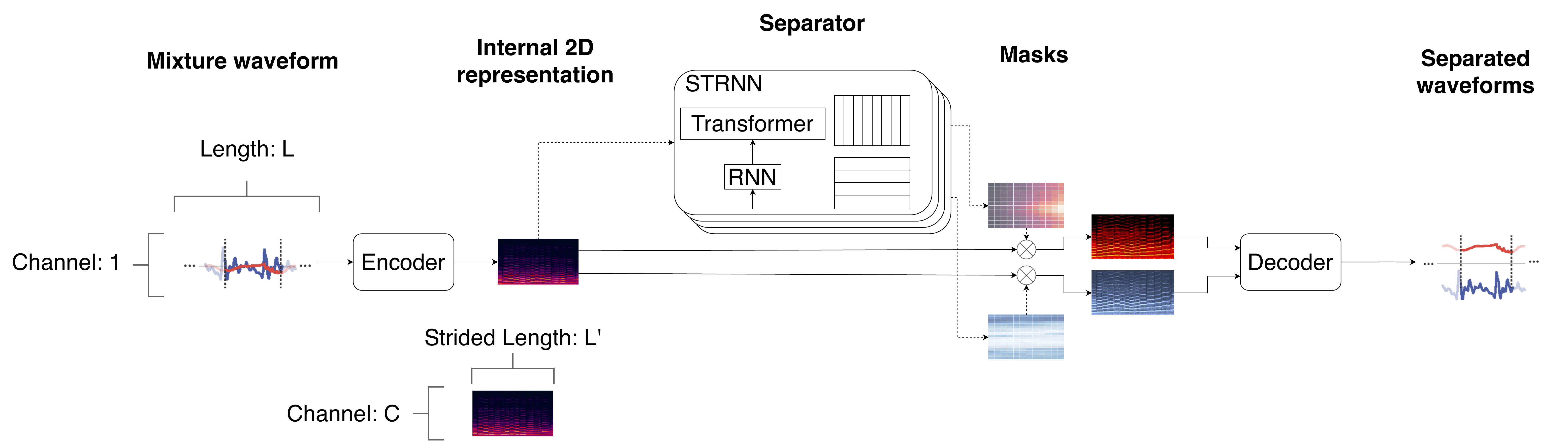}
    \vspace{-5pt}
    \caption{The overall architecture of TransMask: Transformer is run over a group of RNNs in order to better capture both local and global dependency over the time domain.}
    \label{fig:model}
    \vspace{-10pt}
\end{figure*}


 
There are also other applications of transformers.
DPTNet~\cite{chen2020dual} uses similar structure as DPRNN, but replaces the RNN modules with transformers.
Note that the transformers in DPTNet utilize RNN as feed-forward network, which means that DPTNet is still an auto-regressive model. Thus the performance improvement of DPTNet may come from the increased model complexity.
Sepformer~\cite{subakan2020attention} uses pure transformers but the model size is at least 10 times larger than TransMask.
Conformer~\cite{chen2020continuous} adopts the structure which is proved to be successful in speech recognition. This work mainly focuses on the Continuous Speech Separation (CSS) scenario.


\section{Preliminaries}
\label{ssec:bg}



\noindent\textbf{Transformer:} Original transformer~\cite{vaswani2017attention} is mainly used in natural language processing tasks like language model, or neural machine translation. 
 The core component of transformer is the self-attention module followed by a feed forward network. Each element of the sequence attends to all the elements of the same sequence.
 
 Transformer is able to capture the dependency from arbitrary long distance, due to this self-attention mechanism.
 Although it is better than CNN or RNN in capturing long-term dependencies, it may overlook local dependencies, because of the lack of positional information.
 It is therefore important to introduce positional encoding into Transformer to help the model enhance with local dependency information,
 Given positional encoding, it still needs deep transformer layers and plenty of training for the model to fully capture the local dependency.
 To overcome this drawback, some studies, e.g., R-transformer~\cite{wang2019r}, inject RNN modules into transformers; others~\cite{child2019generating, beltagy2020longformer} attend only to a local window or elements from strided distances.
 Different from such methods, we use a strided sparse transformer~\cite{child2019generating} only for handling the long-term dependencies.
 
\noindent\textbf{DPRNN:} DPRNN~\cite{luo2020dual} is a variant of TasNet~\cite{luo2018tasnet}. They share a similar architecture, consisting of an encoder, a separator and a decoder.
The encoder and the decoder can be viewed as trainable substitute of STFT and inversed-STFT as discussed in last section.
DPRNN uses a dual path process for integrating RNN into the separator.
The model first splits the sequence into overlapping chunks, and then performs two paths of RNN. It is done within the chunks first, and across the chunks later.
The input audio with $L$ frames is first padded to be divisible by $P$.
Let $2P$ be the length of the chunk ($P$ is the number of overlapping frames of the two consecutive chunks) and $S$ is the number of chunks.
The intra-chunk RNN is used for capturing the local dependencies. And it is applied on the dimension with length $2P$.
The inter-chunk RNN is used to handle long-term dependencies. It is applied on the dimension with length $S$.
After applying this dual-path RNN multiple times, DPRNN gets a promising result on speech separation tasks.
For DPRNN, the inference efficiency depends on the length of the audio, denoted by $N$. If the chunk size is fixed, the time complexity of DPRNN 's inference is linear to the audio length, i.e., $O(N)$.
 

\section{TransMask}
\label{sec:proposed}

In order to utilize the capability of transformers on handling long-term dependencies, while keeping the auto-regressive regularization which provides important local dependency information, 
TransMask adopts and enhances the ideas of strided sparse transformer and the dual-path process from DPRNN.

\begin{figure}[htb]

\begin{minipage}[b]{0.3\linewidth}
  \centering
  \centerline{\includegraphics[width=2.5cm]{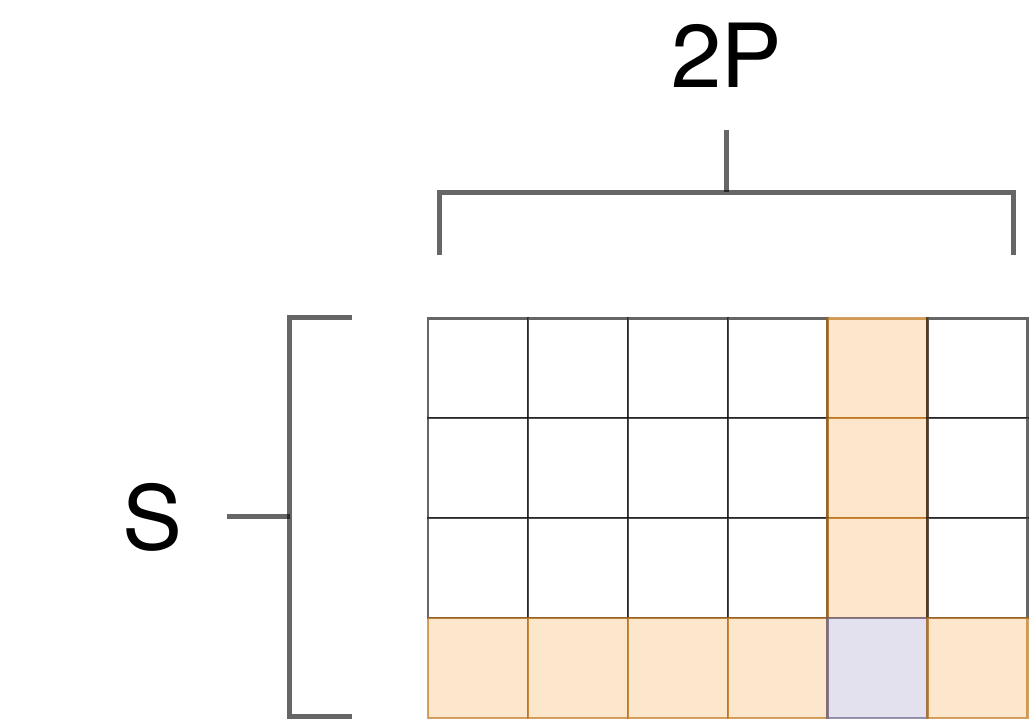}}
  \centerline{(a) DPRNN}\medskip
\end{minipage}
\begin{minipage}[b]{0.3\linewidth}
  \centering
  \centerline{\includegraphics[width=2.5cm]{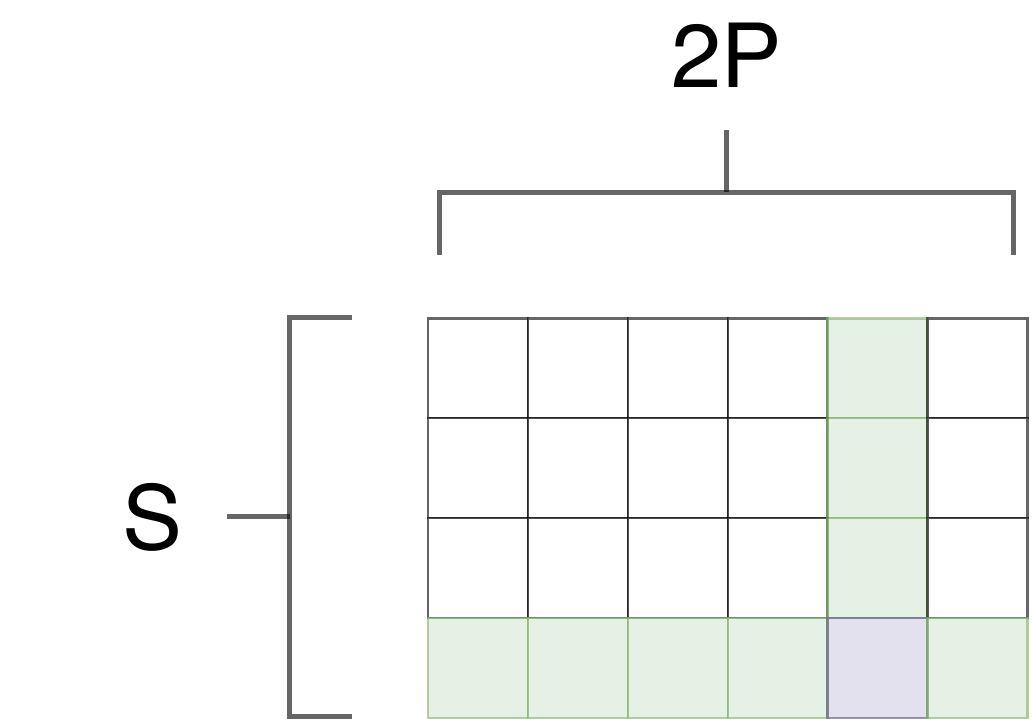}}
  \centerline{(b) Sparse transformer }\medskip
\end{minipage}
\begin{minipage}[b]{0.3\linewidth}
  \centering
  \centerline{\includegraphics[width=2.5cm]{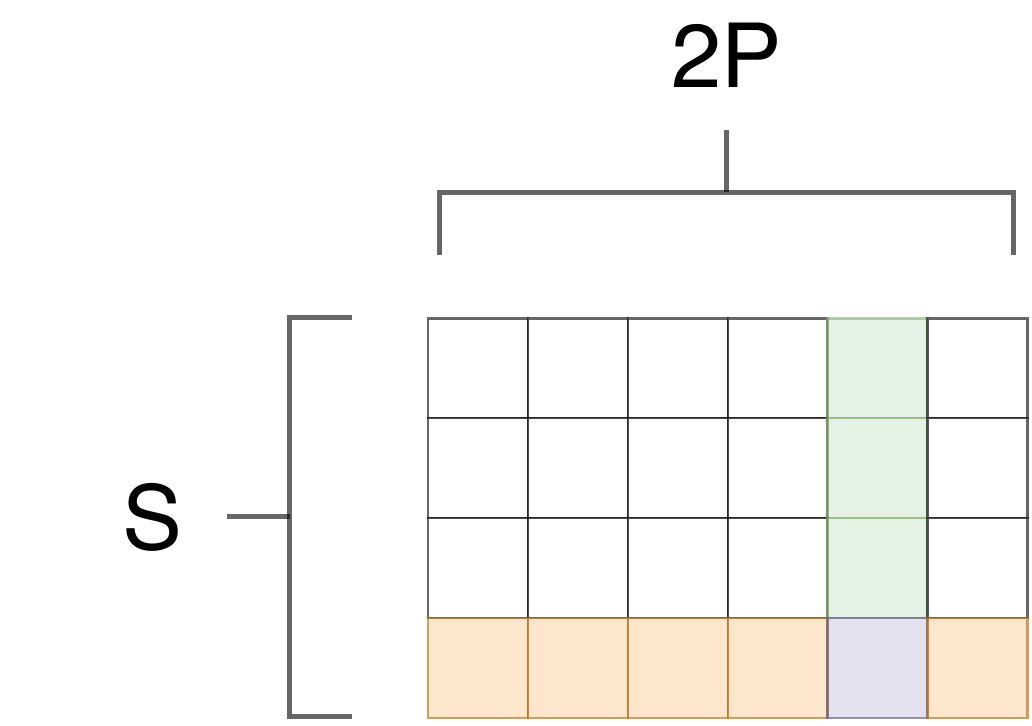}}
  \centerline{(c) STRNN}\medskip
\end{minipage}
\caption{Examples for demonstrating the difference on processing strategies in STRNN of TransMask, DRPNN and Sparse Transformer. }
\label{fig:cmp}
\end{figure}

\noindent\textbf{STRNN: }
The mixture audio first goes through a chunk level process in a bidirectional-LSTM layer, and then it is passed to a strided transformer structure for an inter-chunk attention process.
This architecture connects each frame of the sequence to two kinds of contexts: local context and strided context respectively.
The local context is processed by RNN, and the strided context is handled by the transformer.
We call this structure a strided-transformer with RNN, or STRNN in short.
The difference among STRNN, DPRNN and strided sparse transformer is illustrated in Figure \ref{fig:cmp}.
The grey cell in the graph is the current frame the model is processing.
The orange cells are the frames connected to the current frame by RNN modules.
The green cells are the frames connected to the current frame by self-attention modules.
Figure \ref{fig:cmp}(a) shows that DPRNN connects the local context and strided context to the current frame by RNN modules.
Figure \ref{fig:cmp}(b) shows that strided sparse transformers connect the contexts using only self-attention.
In Figure \ref{fig:cmp}(c), our STRNN strategy works in a different way. It connects the strided contexts using self-attention while using RNN for local contexts.
The overall architecture is shown in Figure \ref{fig:model}, which is similar to DPRNN except that the separator module is replaced by a stack of STRNN layers.
 
Due to transformer's strong capability on capturing the dependencies over the whole sequence, the proposed model is expected to achieve better results than DPRNN while using fewer parameters.
Regarding inference, the self-attention module of STRNN makes the calculation easily paralleled, and the RNN module runs over the sequences with a fixed chunk size.
Therefore, the cost of sequential operations (mainly RNN) of STRNN is O(1), which ensures the promising inference efficiency of TransMask.
 
\begin{figure}[t]
    \centering
    \includegraphics[width=.94\linewidth]{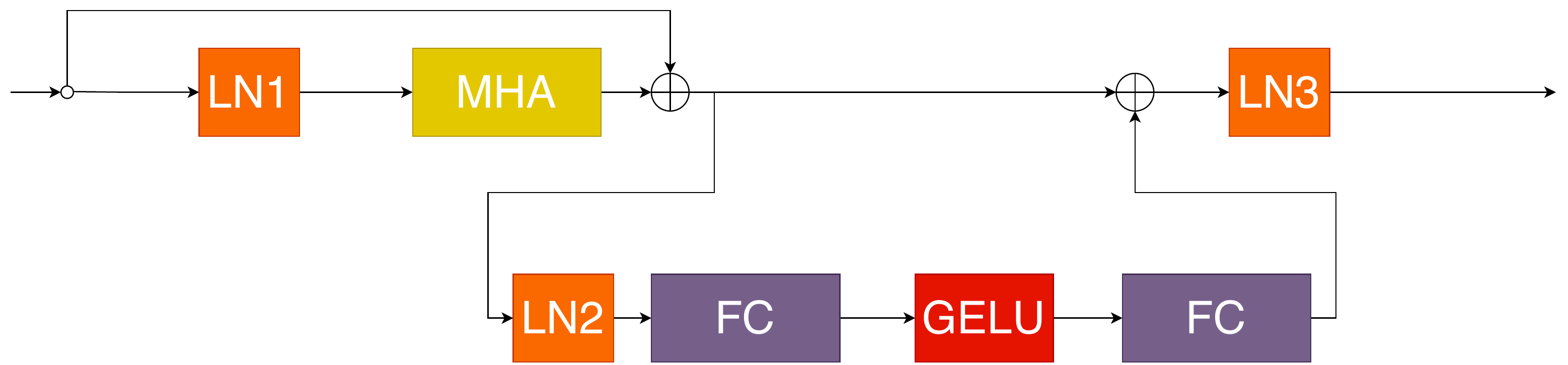}
    \vspace{-5pt}
    \caption{Pre-norm structure used in the proposed model. LN represents layer normalization, MHA is multihead attention, and FC is fully connected layers. We use GELU~\cite{hendrycks2016gaussian} as the activation function.}
    \label{fig:prenorm}
    \vspace{-10pt}
\end{figure}

\noindent\textbf{Sandwich-Norm Transformer Layer: }
To train a good Transformer model, it is important to set up the warm-up steps carefully, where the learning rate  gradually increases.
The number of warm-up steps needs to be tuned, and this makes transformer hard to train.
However, recent work~\cite{nguyen2019transformers} shows that the warm-up may not be necessary, since a change of the order of normalization layers could solve this problem.
When using the pre-norm, warm-up steps become unnecessary.
Our pre-norm transformer layer is shown in Figure \ref{fig:prenorm}. An additional layer normalization is added to the end to avoid the whole transformer layer being by-passed, as ~\cite{wang2020transformer} does. 
We call this a sandwich-norm transformer layer. We empirically found that this structure significantly improves the convergence speed.

\noindent\textbf{Dual-Temporal Convolutional Encoding: }
When using transformers, the self-attention module treats all the keys equally, and this attention process is order-agnostic.
To inject position information into the self-attention process, Transformers concatenate positional encoding into the model inputs. There are different options for positional encoding, as listed below:
1) \noindent\textit{Sinusoid positional encoding:} It uses sinusoid functions for representing different positions in the sequence~\cite{vaswani2017attention}.
    This encoding can be used for either absolute position or relative positions.
2) \noindent\textit{Frame stacking:} instead of using only one frame per position in the input, this method stacks $n$ contextual frames together and creates a new frame from it~\cite{wang2020transformer}. This is mainly used as a relative positional encoding.
3) \noindent\textit{Convolutional encoding:} It uses a convolutional neural network to encode the input~\cite{mohamed2019transformers}. This is similar to frame stacking, since after this encoding, each frame contains the contextual information.
The difference is that this scheme makes the positional encoding trainable. Recent study on speech recognition with transformer proves that convolutional encoding works better than other two methods~\cite{wang2020transformer}. Thus, TransMask chooses the convolutional encoding. 

Different from the 2D CNN modules used in ~\cite{wang2020transformer},
our method does not perform the convolution on the time dimension and the filter bank dimension.
Instead, it splits the input sequence into overlaped chunks, and performs 2D convolution over the intra-chunk dimension and inter-chunk dimension.
Since the 2D convolution is performed on the two temporal dimensions, we call this a dual-temporal convolutional encoding.
Each frame of the positional encoding contains not only the positional information on local contexts, but also the positional information on strided contexts.

\noindent\textbf{Loss Function: }
For the training objectives, we use the SI-SNR with utterance-level Permutation Invariant Training (uPIT) as used in ~\cite{luo2020dual}.

\section{Experiments}
\label{sec:exp}

\noindent\textbf{Dataset:}
Existing studies on speech separation usually use a mixture version of Wall Street Journal (WSJ0), known as WSJ0-2mix ~\cite{hershey2016deep}.
However, WSJ0 only contains 101 different speakers, and 25 hours of training data.
The number of speakers is too small for the evaluation of generalization ability of the models.
LibriMix~\cite{cosentino2020librimix} is recently proposed as a new benchmark using open source dataset LibriSpeech~\cite{panayotov2015librispeech}. It contains 1,172 speakers, and 465 hours of training data.
In this paper, we use the version of Librimix with 2 speakers' mixtures, Libri2Mix. And we only use the part of \emph{train-360} from the LibriSpeech dataset. The codes for mixture generation are available on github~\url{https://github.com/JorisCos/LibriMix}.
 

\noindent\textbf{Model specifications:}
\label{ssec:spec}
We test the proposed model with 4 and 6 layers of STRNN modules, denoted as TransMask-4 and TransMask-6 with learnable time-frequency basis.
The dimension size of the strided transformer is set at 64 for self-attention. The feed forward network has 256 nodes in the hidden layer.
For the convolutional encoding, it contains 3 layers of blocks, each consisting of a convolutional network with kernel size 3, global layer normalization and GELU activation.
The numbers of input channels and output channels are both 64. The output of the separation module is passed to sigmoid activation function. The result is multiplied with the input for getting the learnable basis representation of the clean source predictions.
The code is open-sourced: \url{https://github.com/Speech-AI/SpeechX}. It is based on the asteroid~\cite{Pariente2020Asteroid} project.

\noindent\textbf{Experiment results:}
We compare our model with DPRNN in three aspects: SDR, model size and inference speed.
 As shown in Table \ref{tab:perf}, TransMask-6 achieves the best SDR at 16.3, outperforming DPRNN by a significant margin.
TransMask-4 and TransMask-6 is 50\% and 40\% smaller than DPRNN, respectively.
The inference speed is also at least two times faster.
In the original paper of DPRNN, it claims to achieve $O(\sqrt{N})$ processing time ($N$ stands for the sequence length) if it carefully sets the chunk size and the number of chunks to be close to each other.
However, this can only be controlled during training by fixing the length of the input audio.
When inferencing, the length of the input audio is not controllable. Since the chunk size is a constant, the inference time of DPRNN remains $O(N)$ in practice.
In comparison, our proposed model always uses constant time due to the parallel attribute of self-attention, as long as the inference device has enough resources.
The results are reported in Table \ref{tab:rtf}. They are tested on Ksyun Virtual Senior CPU (ksyun-cpu64-senior). We use real time factor (rtf) as the metric.
We augment the test dataset by simply repeating the audio multiple times.
As the length of the audio goes to 4 times, DPRNN has relatively stable rtf, while rtf of TransMask-6 drop linearly since the inference is in constant time.
The ratio shows that, as the audio lengths increase, the advantage of TransMask gets more significant.
It is even 4 times faster than DPRNN when we expand the test audios 4 times longer.
This phenomenon continues until the CPU resources reaches its limit for the parallel computation, where the audio lengths are expanded to 8 times.

\begin{table}[h!]
    \centering
    \caption{Model size and SDR comparison between previous work and the proposed model}
    \label{tab:perf}
    \begin{tabular}{l|c|c}
        \hline
          & Model size & SDR \\
        \hline
        Conv-TasNet & 5.1M & 13.5 \\
        DPRNN & 2.6M & 15.6 \\
        TransMask-4 & 1.24M & 15.5 \\
        TransMask-6 & 1.62M & 16.3 \\
        \hline
    \end{tabular}
\end{table}

\begin{table}[h!]
    \centering
    \caption{The real time factor (rtf) when we expand the length of the test audio, tested using CPU on Ksyun cloud server}
    \label{tab:rtf}
    \begin{tabular}{l|c|c|c|c}
        \hline
          & original length & 2x & 4x & 8x\\
        \hline
        DPRNN & 1.08 & 0.84 & 0.69 & 0.70 \\
        TransMask-6 & 0.56 & 0.31 & 0.17 & 0.21 \\ 
        \hline
        Ratio & 51.9\% & 36.9\% & 24.6\% & 30.0\% \\
        \hline
    \end{tabular}
\end{table}

%
%
%
%
%

\section{Conclusion}
\label{sec:conclusion}

In this paper, we combine the strength of Transformer and RNN into the architecture for better long-term and short-term dependency handling. We demonstrate that Transformer is effective on reducing the size of the model, improving the inference efficiency, and maintaining or even improving the quality of output speech audio. 

\bibliographystyle{IEEEbib}
\bibliography{strings,refs}

\end{document}